\documentclass[twocolumn,prl,aps,longbibliography,superscriptaddress]{revtex4-1}

\usepackage{times}

\usepackage{amsmath}
\usepackage{amsfonts}
\usepackage{amssymb}
\usepackage{graphicx}
\usepackage{hyperref}
\usepackage{bm}
\usepackage{color}

\usepackage{amstext}
\usepackage{amsxtra}
\usepackage{float}
\usepackage{upgreek}

\hypersetup{backref,
colorlinks=true,
linkcolor=cyan,
linktoc=black,
citecolor=cyan,
urlcolor=cyan}

\renewcommand{\figurename}[1]{Fig. }

\begin{document}

\title{Universal Prethermal Dynamics of Bose Gases Quenched to Unitarity}
\author{Christoph Eigen}
\affiliation{Cavendish Laboratory, University of Cambridge, J. J. Thomson Avenue, Cambridge CB3 0HE, United Kingdom }
\author{Jake A.~P. Glidden}
\affiliation{Cavendish Laboratory, University of Cambridge, J. J. Thomson Avenue, Cambridge CB3 0HE, United Kingdom }
\author{Raphael Lopes}
\altaffiliation[Present address: ]{Laboratoire Kastler Brossel, Coll{\`e}ge de France, CNRS, ENS-PSL University,
UPMC-Sorbonne Universit{\'e}, 11 Place Marcelin Berthelot, F-75005 Paris, France}
\affiliation{Cavendish Laboratory, University of Cambridge, J. J. Thomson Avenue, Cambridge CB3 0HE, United Kingdom }
\author{Eric A. Cornell}
\affiliation{JILA, National Institute of Standards and Technology and University of Colorado, and Department of Physics, Boulder, Colorado 80309-0440, USA}
\author{Robert P. Smith}
\affiliation{Cavendish Laboratory, University of Cambridge, J. J. Thomson Avenue, Cambridge CB3 0HE, United Kingdom }
\affiliation{Clarendon Laboratory, University of Oxford, Oxford OX1 3PU, United Kingdom}
\author{Zoran Hadzibabic}
\affiliation{Cavendish Laboratory, University of Cambridge, J. J. Thomson Avenue, Cambridge CB3 0HE, United Kingdom }




\maketitle

{\bf 
Understanding strongly correlated phases of matter, from the quark-gluon plasma to neutron stars, and in particular the dynamics of such systems, {\it e.g.} following a Hamiltonian quench, poses a fundamental challenge in modern physics. Ultracold atomic gases are excellent quantum simulators for these problems, thanks to tuneable interparticle interactions and experimentally resolvable intrinsic timescales. In particular, they give access to the unitary regime where the  interactions are as strong as allowed by quantum mechanics. Following years of experiments on unitary Fermi gases ~\cite{Zwerger:2011,Zwierlein:2014}, unitary Bose gases have recently emerged as a new experimental frontier~\cite{Navon:2011,Rem:2013,Fletcher:2013,Makotyn:2014,Eismann:2016,Fletcher:2017,Klauss:2017,Eigen:2017,Fletcher:2018}. They promise exciting new possibilities~\cite{Chevy:2016}, including universal physics solely controlled by the gas density~\cite{Cowell:2002,Ho:2004a} and novel forms of superfluidity~\cite{Radzihovsky:2004,Romans:2004,Piatecki:2014}. Here, through momentum- and time-resolved studies, we explore both degenerate and thermal homogeneous Bose gases quenched to unitarity. In degenerate samples we observe universal post-quench dynamics in agreement with the emergence of a prethermal state~\cite{Berges:2004,Gring:2012,Yin:2013,Sykes:2014,Kain:2014,Rancon:2014,Yin:2016} with a universal nonzero condensed fraction~\cite{Kain:2014,Yin:2016}. In thermal gases, dynamic and thermodynamic properties generically depend on both the gas density $n$ and temperature $T$, but we find that they can still be expressed in terms of universal dimensionless functions. Surprisingly, the total quench-induced correlation energy is independent of the gas temperature. Our measurements provide quantitative benchmarks and new challenges for theoretical understanding. 
}

\begin{figure}[t]
\centering
\includegraphics[width=\columnwidth]{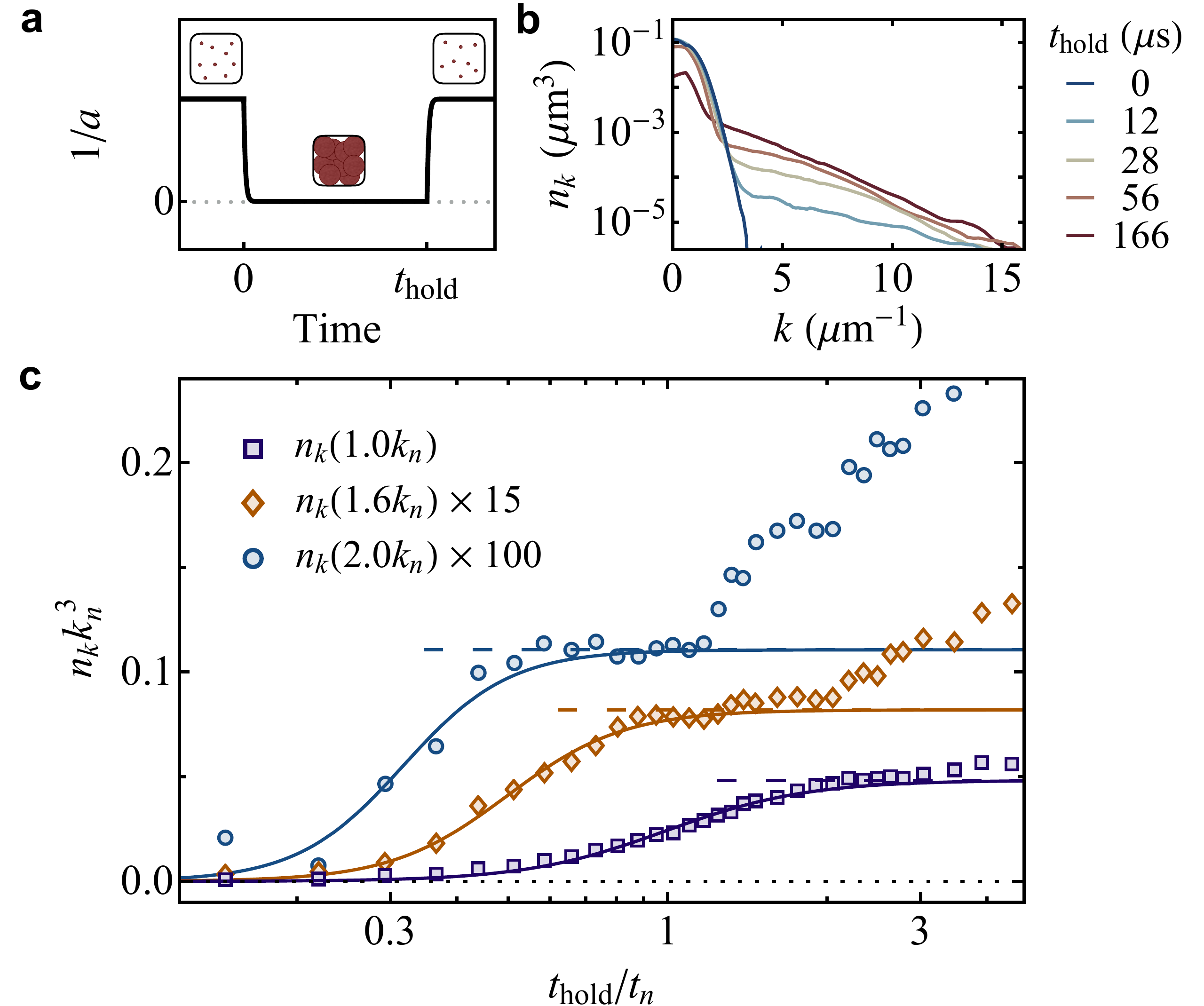}
\caption{Dynamics of a degenerate Bose gas quenched to unitarity. 
(a) Quench protocol. The red circles depict atoms, and their sizes the interaction strength, limited at unitarity by the interparticle spacing. 
(b) Momentum distributions for various $t_{\rm{hold}}$; here $n= 5.1~\mu{\rm m}^{-3}$, so $k_n = 6.7~\mu {\rm m}^{-1}$ and $t_n = 27~\mu$s.
(c) Populations of individual $k$ states show a rapid initial growth, saturation at (quasi-)steady-state values $\overline{n_k}$ (dashed lines), and long-time heating. 
The solid lines show sigmoid fits used to extract the initial-growth times $\tau(k)$.}
\vspace{-3mm}
\label{fig:1}
\end{figure}

\begin{figure*}[tbp]
\includegraphics[width=\textwidth]{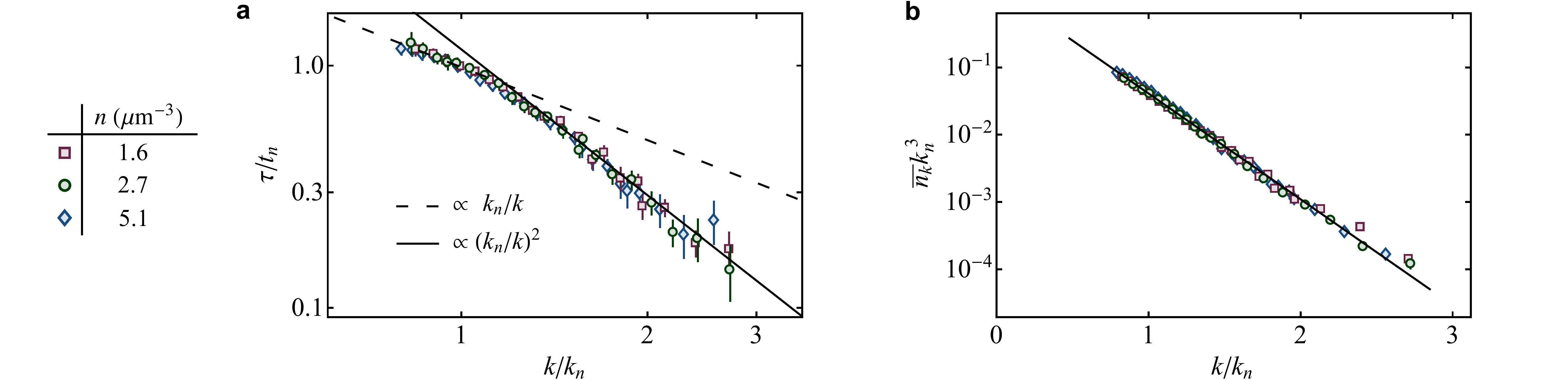}
\caption{Universal post-quench dynamics and the steady-state momentum distribution in the degenerate Bose gas. 
We show (a) $\tau(k)$, and (b) the steady-state $\overline{n_k}(k)$, for three different BEC densities.  Expressing all quantities in dimensionless form, using the natural Fermi scales $t_n$ and $k_n$, collapses all our data onto universal curves.
The solid line in (b) is an exponential fit, $\overline{n_k} k_n^3 = 1.53 \exp[-3.62 \, k/k_n ]$.}
\vspace{-1mm}
\label{fig:2}
\end{figure*}


In ultracold atomic gases two-body contact interactions are characterised by the $s$-wave scattering length $a$, and the unitary regime is realised by tuning $a\rightarrow \infty$, using magnetic Feshbach resonances~\cite{Chin:2010}. In Bose gases this also enhances three-body recombination that leads to particle loss and heating, making them inherently dynamical, non-equilibrium systems. Experimentally, they are studied by rapidly quenching $a\rightarrow \infty$ (see Fig.~\ref{fig:1}(a)), which initiates the non-equilibrium dynamics.  If one starts with a Bose-Einstein condensate (BEC), in the $k \approx 0$ momentum state, after the quench the momentum distribution broadens (kinetic energy grows), both due to lossless correlation dynamics and due to recombination heating; see Fig.~\ref{fig:1}(b). The interplay of these processes raises many challenging questions. Eventually, the condensate inevitably vanishes, but does the gas attain a strongly-correlated  quasi-equilibrium steady state before degeneracy is lost?  
If so, what is the nature of this state?

The timescales for the different processes are set by the natural lengthscales in the system. Within the universality hypothesis~\cite{Ho:2004a}, in a homogeneous degenerate unitary gas the only relevant lengthscale is the interparticle spacing $n^{-1/3}$, which (in analogy with Fermi gases) sets the Fermi momentum $\hbar k_n = \hbar (6\pi^2 n)^{1/3}$, energy $E_n = \hbar^2 k_n^2/(2m)$, and time $t_n = \hbar/E_n$; here $m$ is the particle mass. Additional potentially relevant lengthscales are the sizes of the Efimov trimer states that exist due to resonant two-body interactions~\cite{Efimov:1970,Kraemer:2006, Smith:2014, Piatecki:2014, Comparin:2016,Colussi:2018,DIncao:2018b}. 
Experimentally, three-body correlations~\cite{Fletcher:2017} and Efimov trimers~\cite{Klauss:2017} have been observed, but all degenerate-gas dynamics have been consistent with $t_n$ being the only characteristic timescale~\cite{Makotyn:2014,Klauss:2017,Eigen:2017}. This fascinating universality also has a downside - it has so far made it impossible to disentangle the lossless and the recombination-induced dynamics. First experimental evidence suggested that the lossless processes are somewhat faster, sufficiently so that the gas attains a degenerate steady state~\cite{Makotyn:2014,Eigen:2017}, but almost nothing could be established about its nature. Here we isolate the effects of the lossless post-quench dynamics through momentum- and time-resolved studies of both degenerate and thermal Bose gases.

We prepare a homogeneous $^{39}$K Bose gas in an optical-box trap of volume $\sim~3 \times 10^4~\mu$m$^3$~\cite{Gaunt:2013,Eigen:2017}, and use a Feshbach resonance centred at $402.70(3)$~G~\cite{Fletcher:2017}. Initially we prepare either a quasi-pure BEC or a thermal gas. In both cases we start with a weakly interacting sample, with $n a^3<10^{-4}$, then quench $a\rightarrow \infty$ (within $2~\mu$s), let the gas evolve for a time $t_{\rm{hold}}$, quench back to low $a$, and finally release the gas and measure its momentum distribution $n_k(k)$; see Methods for technical 
details. We normalise $n_k$ so that $\int 4\pi k^2 n_k \, {\rm d}k=1$.

We first present our study of degenerate gases.
In Fig.~\ref{fig:1}(b) we show $n_k (k)$ for initial BEC density $n= 5.1~\mu{\rm m}^{-3}$ and various $t_{\rm hold}$. 
In Fig.~\ref{fig:1}(c) we illustrate our key experimental observation: looking at  $n_k$ values for individual $k$ states, we do discern separate stages in their evolution -  after a rapid initial growth, $n_k$ approaches a steady-state value $\overline{n_k}$ (dashed lines), before the long-time heating takes over. All timescales are $\sim t_n$, but still distinguishable.  We discern such time-separation for all  $k/k_n \gtrsim 0.8$.
Using sigmoid fits (solid lines) we extract the time $\tau(k)$ for the initial rapid growth of $n_k$, defined such that $n_k(k, \tau(k)) = \overline{n_k}(k)/2$. Note that here, and throughout the paper, $t_n$ and $k_n$ correspond to the initial $n$; for our longest $\tau$ we observe $\approx 20\%$ particle loss.


Crucially, in Fig.~\ref{fig:1}(c) we also see that the different-$k$ curves are not aligned in time; $n_k (2k_n)$ shows signs of heating before $n_k(k_n)$ reaches its steady-state value. This illustrates why one could not quantitatively separate lossless and recombination dynamics by considering all $k$'s at the same evolution time~\cite{Makotyn:2014,Eigen:2017}, for example by looking at the kinetic energy per particle, $E(t_{\rm hold})$~\cite{Eigen:2017}. Instead, we separately obtain $\overline{n_k}$ values for different $k$ and piece together the function $\overline{n_k}(k)$. 
This {\it does not} give the momentum distribution at any specific time, but instead it allows us to infer  what the steady-state $n_k(k)$ would be if the gas did not suffer from losses and heating. 
Here we assume that at early times, $t_{\rm hold} \sim t_n$, all nonzero-$k$ states are primarily fed from the macroscopically occupied BEC (see Fig.~\ref{fig:1}(b)).

In Fig.~\ref{fig:2} we plot the dimensionless $\tau/t_n$ and $\overline{n_k} k_n^3$ versus the dimensionless $k/k_n$, for three different BEC densities. We see that by expressing all quantities in such dimensionless form, all our data points fall onto universal curves.

The universal $\tau/t_n$ is $\propto k_n/k$ at low $k$ and $\propto (k_n/k)^2$ at high $k$. This was qualitatively predicted for the emergence of a prethermal steady state~\cite{Yin:2013, Sykes:2014, Kain:2014, Rancon:2014,Yin:2016}. In this picture, at short times after the quench, the excitations are similar to the Bogoliubov modes in a weakly interacting BEC, phonons at low $k$ and particles at high $k$, but with the usual mean-field energy replaced by an energy $\sim E_n$. The speed of sound is then $\sim \hbar k_n/m$ and the crossover between the two regimes is at $k \sim k_n$. Finally, $\tau(k)$ is set by the dephasing time, given approximately by the inverse of the excitation energy.

\begin{figure*}[t]
\centering
\includegraphics[width=\textwidth]{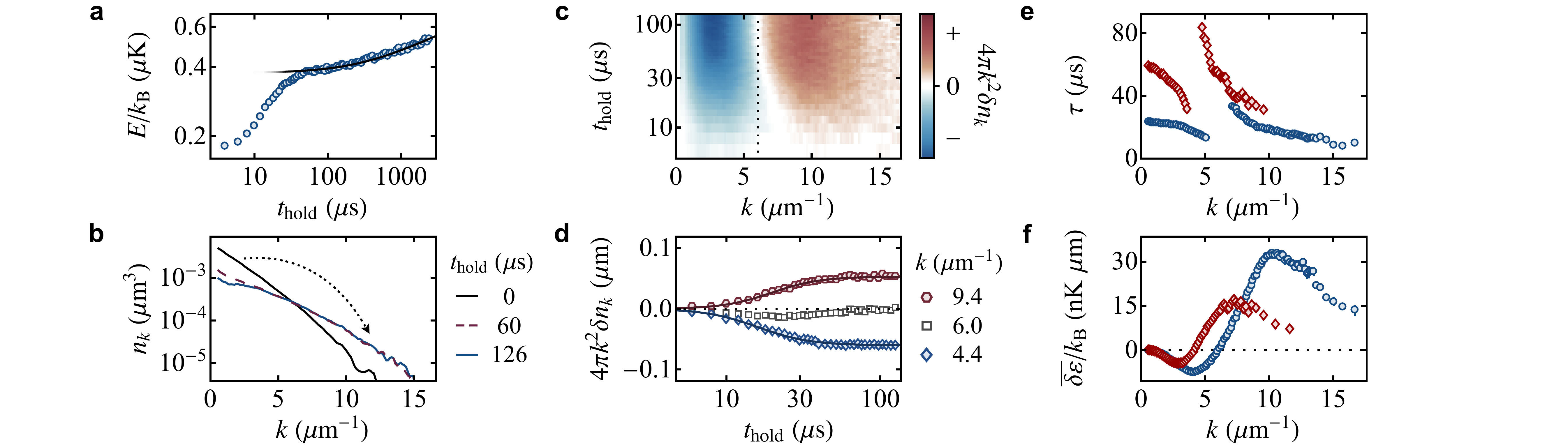}
\caption{Thermal Bose gas quenched to unitarity.
(a) The kinetic energy per particle, $E$, shows a rapid growth at $t_{\rm hold} \lesssim 100~\mu$s, and significant heating only for $t_{\rm hold} \gg 100~\mu$s; the black line is the prediction for recombination heating. Here, and in (b-d), $n= 5.6~\mu$m$^{-3}$ and $T = 150$~nK.
(b) Momentum distributions at different times after the quench. The initial redistribution of particles from low to high $k$ is essentially complete within $60~\mu$s, and $n_k(k)$ is almost identical at $126~\mu$s. 
(c) Population changes in different $k$-space shells; the population in $k_0 = 6.0~\mu{\rm m}^{-1}$ (dashed line) remains essentially unchanged. 
(d) Vertical cuts through the plot in (c). Solid lines are sigmoid fits used to extract $\tau(k)$. 
(e, f) $\tau(k)$ and $\overline{\delta \varepsilon}(k) \propto k^4 \, \overline{\delta n_k}(k)$, for $n= 5.6~\mu$m$^{-3}$ and $T = 150$~nK (blue), and for $n= 1.3~\mu$m$^{-3}$ and $T = 70$~nK (red).}
\vspace{-1mm}
\label{fig:3}
\end{figure*}

The universal $\overline{n_k} k_n^3$ curve is more surprising and poses a new theoretical challenge. Empirically, over three decades in $\overline{n_k}k_n^3$, our data is captured well by a simple exponential, $A \exp[-B k/k_n ]$, with $A=1.53(5)$ and $B=3.62(2)$ (see also Methods). Taken at face value, this function implies a condensed fraction of $\eta  = 1 - \int 4\pi k^2 \overline{n_k} \, {\rm d} k = 19(4)\%$.  
Up to $k\approx 3k_n$ we do not observe the asymptotic form $n_k \sim 1/k^4$ expected at very high $k$~\cite{Tan:2008}, but even if $n_k$ changed to this slower-decaying form right outside of our experimental range, $\eta$ would change by $<3\%$. Our estimate of $\eta$ is close to the predictions for a prethermal state in Refs.~\cite{Kain:2014,Yin:2016}, but the exponential $\overline{n_k} k_n^3$ has (to our knowledge) not been theoretically predicted. Explaining this experimental observation may require explicitly considering the quench away from unitarity.


We now turn to thermal gases, which reveal some simplifications, but also more surprises. A simplification is that in a thermal gas three-body recombination is slowed down more than the lossless dynamics~\cite{Li:2012,Rem:2013,Fletcher:2013}. As shown in Fig.~\ref{fig:3}(a), now simply looking at $E(t_{\rm hold})$ one clearly sees two separate stages in the post-quench dynamics - a rapid initial growth and long-time heating. The shape of the curve is similar to those seen for individual $k$ states in Fig.~\ref{fig:1}(c), and the long-time energy growth matches the theory of recombination heating~\cite{Rem:2013,Eigen:2017}. All this reinforces our interpretation of the two-step dynamics, for both degenerate and thermal gases. We now focus on the early-time dynamics, at $t_{\rm hold} \lesssim 100~\mu$s in Fig.~\ref{fig:3}(a). As we show in Fig.~\ref{fig:3}(b), $n_k(k)$ is essentially identical at $60$ and $126~\mu$s, meaning that on this timescale a steady state is established  for all $k$.

In a thermal gas, even before the quench to unitarity $n_k$ is significant for $k \lesssim 1/\lambda$, where $\lambda = h/\sqrt{2\pi m k_{\rm B} T}$ is the thermal wavelength (here, and everywhere below, $T$ is the initial temperature, before the quench to unitarity). We thus look at the redistribution of particles in $k$-space - the change $\delta n_k (k)$ with respect to $t_{\rm hold} =0$, and the corresponding change in the spectral energy density $\varepsilon  = \hbar^2/(2m) \, 4\pi k^4 n_k$.  A new challenge is that we now have two relevant lengthscales, $n^{-1/3}$ and $\lambda$,  and it is not {\it a priori} clear if the dynamic and thermodynamic properties can be expressed in terms of dimensionless universal functions.

Fig.~\ref{fig:3}(c) shows time-resolved population changes in different spherical shells in $k$-space, $4\pi k^2 \delta n_k$. We see that for some special $k_0$ (dashed line) the population remains essentially constant. Fig.~\ref{fig:3}(d) shows vertical cuts through Fig.~\ref{fig:3}(c) for $k<k_0$, $k=k_0$, and $k>k_0$. Away from $k_0$, we use sigmoid fits (solid lines) to extract $\tau(k)$, now for both diminishing and growing populations. Near $k_0$ we see just a small wiggle in $\delta n_k$, to which we cannot assign a single timescale.

In Fig.~\ref{fig:3}(e,f) we show $\tau(k)$ and the steady-state $\overline{\delta \varepsilon}(k)$ for two different combinations of $n$ and $T$. The $\overline{ \delta \varepsilon}(k)$ curve intuitively conveys the redistribution of particles from $k<k_0$ to $k>k_0$, and the resulting energy growth $\overline{\Delta E} = \int \overline{\delta \varepsilon} \, {\rm d} k$. The dispersive shape of $\tau (k)$ was not anticipated and invites further theoretical work. Here we empirically ask whether these curves can be scaled into universal dimensionless functions.

\begin{figure*} [t]
\centering
\includegraphics[width=\textwidth]{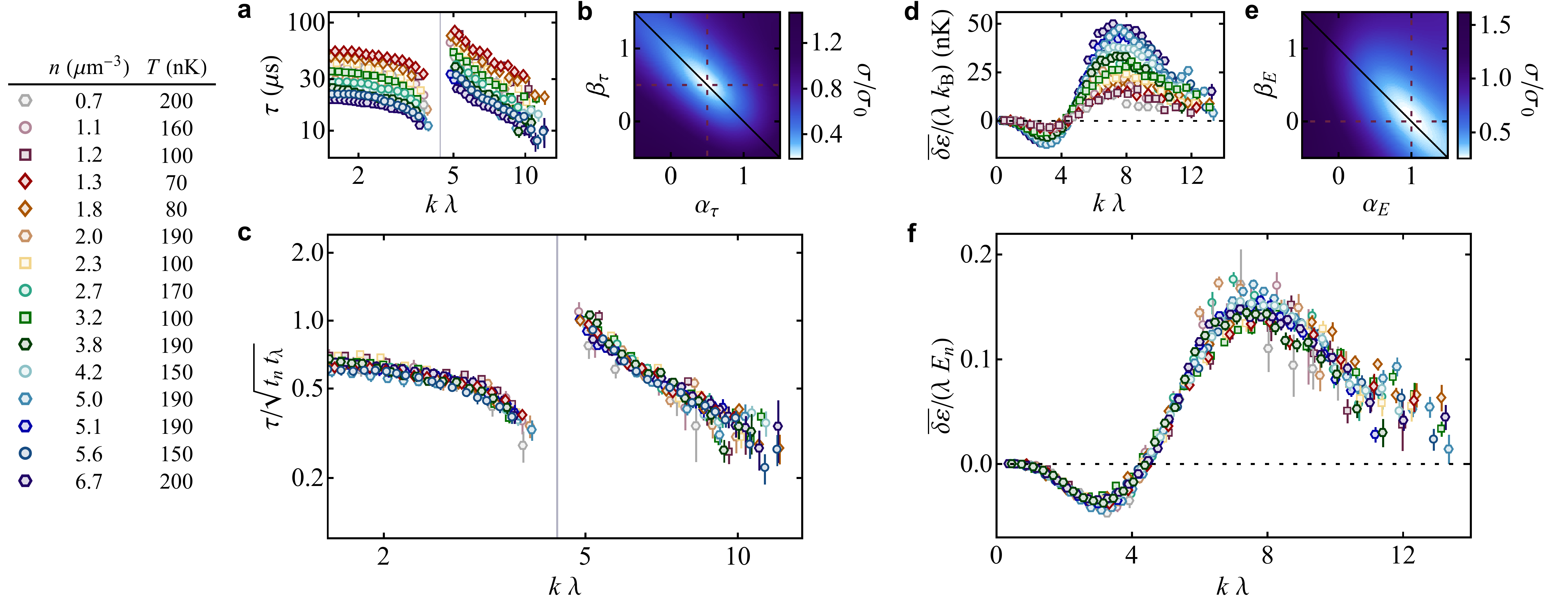}
\caption{Universal dynamic and thermodynamic functions for the thermal Bose gas quenched to unitarity.
(a, d) Plotting $\tau$ and $\overline{\delta \varepsilon}/\lambda$ versus $k\lambda$ horizontally aligns all our curves for 15 different combinations of $n$ and $T$. 
(b, e) Supposing that the characteristic timescale for the dynamics is $t_s \sim t_n^{\alpha_t} t_{\lambda}^{\beta_t}$, we get the best data collapse for $\alpha_t \approx \beta_t \approx 1/2$, suggesting $t_s = \sqrt{t_n t_{\lambda}}$ (see the text for details). Similarly, for the energy scale $E_s \sim E_n^{\alpha_E} (k_{\rm B} T)^{\beta_E}$ we get $\alpha_E \approx 1$ and $\beta_E \approx 0$, implying $E_s = E_n$.
(c, f) The dimensionless $\tau/ \sqrt{t_n t_{\lambda}}$ and $\overline{\delta \varepsilon}/(\lambda E_n)$ are, to within experimental errors, universal functions of the dimensionless $k \lambda$.}
\vspace{-0mm}
\label{fig:4}
\end{figure*}

For the horizontal scaling we find that the natural scale for $k$ is $1/\lambda$, independent of $n$. In Fig.~\ref{fig:4}(a) we plot $\tau(k)$ versus $k \lambda$, now for $15$ combinations of $n$ and $T$ (corresponding to phase-space density $n\lambda^3$ between $0.2$ and $2$). Similarly, in Fig.~\ref{fig:4}(d) we plot $\overline{\delta \varepsilon} (k)/\lambda$ versus $k\lambda$, so that the area under each curve is still  $\overline{\Delta E}(n,T)$. In both cases we see horizontal alignment of all the curves, with $k_0 = 4.4/\lambda$.

A more challenging question is whether these $n$- and $T$-dependent curves can be collapsed vertically, by scaling them with some time $t_s(n,T)$ and energy $E_s(n,T)$. To this we take a heuristic approach. We conjecture that  $t_s \sim t_n^{\alpha_t} t_{\lambda}^{\beta_t}$, where $t_{\lambda} = \hbar/(k_{\rm B} T)$, and similarly $E_s \sim E_n^{\alpha_E} (k_{\rm B} T)^{\beta_E}$, and ask for which $\alpha$ and $\beta$ values we get the best collapse. We treat $\alpha$ and $\beta$ exponents as independent, but physically (if there are no other relevant scales) we expect $\alpha_t + \beta_t = \alpha_E + \beta_E=1$.

We quantify the degree of the data collapse by a single number $\sigma$, obtained by calculating the standard deviation of the data points for all $n$ and $T$ at a fixed $k\lambda$ and then summing over $k\lambda$. In Figs.~\ref{fig:4}(b) and (e), respectively, we show plots of $\sigma/\sigma_0$ for $\tau$ and $\overline{\delta \varepsilon}/\lambda$; here $\sigma_0$ corresponds to no scaling. 

For the temporal scaling, in Fig.~\ref{fig:4}(b) we find the lowest $\sigma$ near $\alpha_t = \beta_t = 1/2$, suggesting $t_s = \sqrt{t_n t_{\lambda}}$. 
In Fig.~\ref{fig:4}(c) we plot $\tau/\sqrt{t_n t_{\lambda}}$ and see that all our data collapse onto a universal curve (within experimental scatter).
For this scaling we also have an intuitive interpretation: in a thermal gas particles do not overlap, and to start feeling the unitary interactions after the quench they must meet; up to a numerical factor, $\sqrt{t_n t_{\lambda}} \sim n^{-1/3} \lambda \, m/\hbar$ is their `meeting time', the ratio of the interparticle distance and the thermal speed.

More surprisingly, in Fig.~\ref{fig:4}(e) we find optimal $\alpha_E \approx 1$ and $\beta_E \approx 0$, suggesting that $E_s$  is simply $E_n$. In Fig.~\ref{fig:4}(f) we see that this scaling collapses all our data onto a universal curve. It also, rather remarkably, implies that while $\overline{\delta \varepsilon} (k)$ depends on both $n$ and $T$, its integral $\overline{\Delta E}$ is independent of $T$. 

This lack of $T$-dependence suggests that $\overline{\Delta E} / E_n$ should also be equal to $\overline{E}/E_n$ for a degenerate gas (where $\overline{\Delta E} = \overline{E}$). Bearing in mind the caveat that we do not experimentally see very high-$k$ tails, from the data in Fig.~\ref{fig:4}(e) we estimate $\overline{\Delta E}/E_n = 0.7(1)$ for a thermal gas, and from the exponential $\overline{n_k}k_n^3$ in Fig.~\ref{fig:2}(b) we indeed get a consistent $\overline{E}/E_n =0.74(4)$ for a degenerate gas.

Our experiments establish a comprehensive view of the prethermal dynamics and thermodynamics of homogeneous Bose gases quenched to unitarity, at both low and high temperatures. They provide both quantitative benchmarks and new conceptual puzzles for the theory. Open problems include explaining the forms of our experimentally observed universal dynamic and thermodynamic functions, and elucidating the connections between these universal features and the previously observed signatures~\cite{Fletcher:2017,Klauss:2017} of the non-universal  Efimov physics. Experimentally, a major future challenge is to probe the coherence and the potential superfluid properties of the prethermal state of a degenerate unitary Bose gas.



%


\vspace{1mm}

{\bf Acknowledgments} We thank Richard Fletcher, Nir Navon and Timon Hilker for  discussions and comments on the manuscript. This work was supported by the Royal Society, EPSRC [Grants No. EP/N011759/1 and No. EP/P009565/1], ERC (QBox), AFOSR, and ARO. R.L. acknowledges support from the E.U. Marie-Curie program [Grant No. MSCA-IF-2015 704832] and Churchill College, Cambridge. E.A.C. acknowledges hospitality and support from Trinity College, Cambridge.

\vspace{-0mm}

\section{Methods}

\setcounter{section}{0}
\setcounter{subsection}{0}
\setcounter{figure}{0}
\renewcommand{\figurename}[1]{Extended Data Fig. }
\makeatletter
\renewcommand{\thefigure}{\@arabic\c@figure} 
\makeatother
\renewcommand\thetable{S\arabic{table}}
\renewcommand{\vec}[1]{{\boldsymbol{#1}}}

{\bf Technical details.}
As described in  Refs.~\cite{Gaunt:2013,Eigen:2016}, our box trap is formed by blue-detuned, $532$-nm laser beams, and is of cylindrical shape, with a diameter of about $30~\mu$m and a length of about $45~\mu$m. We deduce $n$ from the measured atom number, and take into account the fact that the trap walls are not infinitely steep~\cite{Gaunt:2013}, due to the diffraction limit on the sharpness of the laser beams, so the effective trap volume depends slightly on the energy per particle in the initially prepared sample.

Our clouds are in the lowest hyperfine ground state and we initially prepare them at a field of $\approx 399.1$~G. At this field the scattering length is $a_i \approx 400\,a_0$, where $a_0$ is the Bohr radius. 

At the end of $t_{\rm hold}$ we quench $a$ back to $a_i$ using an exponential field ramp with a time constant of $1~\mu$s. 
We use our fastest technically possible ramp in order to minimise conversion of atoms into molecules~\cite{Klauss:2017,Eigen:2017}. We then release the gas from the trap and simultaneously (within $\approx 3$~ms) completely turn off interactions ($a\rightarrow 0$). After letting it expand for $6-12$~ms of time-of-fight (ToF), we take an absorption image of the cloud. We typically repeat each such measurement about 20 times. To reconstruct $n_k(k)$ from the two-dimensional absorption images, which give the momentum distribution integrated along the line of sight, we average each image azimuthally, then average over the experimental repetitions, and finally perform the inverse-Abel transform.  
Due to the initial cloud size and non-infinite ToF, our measurements of $n_k(k)$ are not quantitatively reliable for $k < 2~\mu{\rm m}^{-1}$.

\begin{figure}[b]
\centering
\includegraphics[width=\columnwidth]{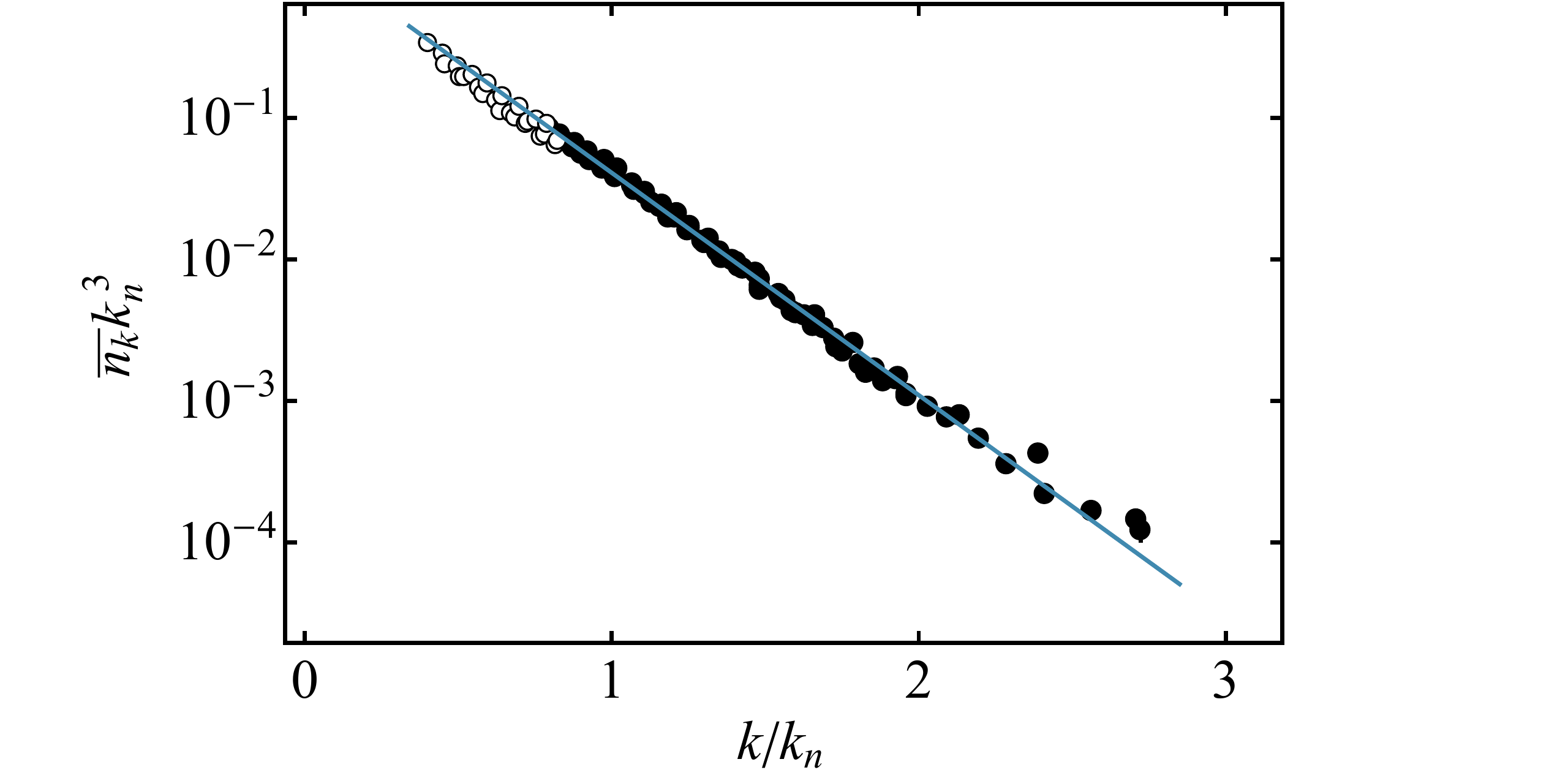}
\caption{Extrapolation of $\overline{n_k} k_n^3$ in a degenerate gas to lower $k/k_n$. Solid symbols: directly measured values also shown in Fig.~\ref{fig:2}(b), here combining the data for all three BEC densities. Open symbols: experimentally extrapolated values, for all three densities, as described in the Methods text. The solid line is the same as in Fig.~\ref{fig:2}(b).
}
\label{fig:M1}
\end{figure}

\vspace{0mm}

{\bf Extrapolation of $\overline{n_k} k_n^3$ in a degenerate gas.}
We can also use our experimental data to estimate how the function $\overline{n_k}k_n^3$ extrapolates to lower $k/k_n$, without presuming its functional form. At $k/k_n< 0.8$ we do not see clear steady-state plateaux in $n_k(t_{\rm hold})$, such as indicated by the dashed lines in Fig.~\ref{fig:1}(c). However, we can extrapolate $\tau \propto t_n k_n/k$, according to the dashed line in Fig.~\ref{fig:2}(a); then, assuming that heating is not yet significant at $t_{\rm hold} = \tau(k)$ and following our definition of $\tau$, we estimate $\overline{n_k} = 2 n_k (\tau)$. Here $n_k (\tau)$ is the measured $n_k$ at the extrapolated $\tau$. These extrapolated values of $\overline{n_k} k_n^3$ are shown by open symbols in Extended Data Fig.~\ref{fig:M1}. They fall on the same exponential curve that fits our directly measured values of $\overline{n_k} k_n^3$ (solid symbols), lending further support for this unexpected functional form.

\end{document}